\documentclass[showpacs,showkeys,twocolumn]{revtex4}%
\usepackage[colorlinks=true]{hyperref}

\AtBeginDocument{
	\hypersetup{
		urlcolor=cyan,
		citecolor=magenta,
		linkcolor=orange,
		anchorcolor=green,
	}%
}
\usepackage{amssymb}
\usepackage{amsmath}
\usepackage{amsfonts}
\usepackage{graphicx}
\usepackage{ulem} 

\usepackage{cancel}
\usepackage[usenames]{xcolor}
\usepackage{epstopdf}
\usepackage{extarrows}
\providecommand{\av}[1]{\left\langle #1 \right\rangle} %
\providecommand{\rbr}[1]{\left( #1 \right)}%

\begin{document}

\title[ ]{Overcoming the Artificial Biases for the Nonadditive $ q $-Entropy}
\author{$^{1}$Thomas Oikonomou}
\email{thomas.oikonomou@nu.edu.kz}
\author{$^{2}$G. Baris Bagci}
\affiliation{$^{1}$Department of Physics, School of Science and
Technology, Nazarbayev University, Astana
010000, Kazakhstan}
\affiliation{$^{2}$Department of Physics, Mersin University, 33110 Mersin, Turkey}

\begin{abstract}

Entropy maximization procedure has been a general practice in many diverse fields of science to obtain the concomitant probability distributions. The consistent use of the maximization procedure on the other hand requires the probability distributions to obey the probability multiplication rule for independent events. However, despite that the nonadditive $ q $-entropy is known not to obey this rule, it is still used with the entropy maximization procedure to infer the probability distributions at the expense of creating artificial biases not present in the data itself. Here we show that this important obstacle can be overcome by considering the intrinsic discrete structure and related averaging scheme of the nonadditive $ q $-entropy. This also paves the road to a better understanding of the entropy maximization procedure of Jaynes.         

\end{abstract}

\eid{ }
\date{\today }
\startpage{1}
\endpage{1}
\maketitle


The concept of entropy is widely used in different contexts such as statistical mechanics \cite{S1,S2,S3,S4,S5}, quantum foundations and information theory \cite{Q1,Q2,Q3,Q4,Q5,Q6}, complex networks \cite{N1,N2,N3}, atmospheric sciences \cite{W1,W2} and biology \cite{Oik1}. In all these different fields, one usually employs the so called Boltzmann-Gibbs-Shannon (BGS) entropy (or von Neumann entropy in quantum mechanics), since its use is foundationally justified in a rigorous manner through for example the Khincin axioms \cite{K}. However, the justification of the entropic form is not sufficient \textit{per se}, since one also practically needs the explicit form of the probability distribution associated with the entropic form. The equilibrium statistical physics in fact succeeded in obtaining this distribution and dubbed it as the canonical distribution. Later, Jaynes obtained the same distribution through an information theoretic analysis now called the entropy maximization procedure i.e. the MaxEnt \cite{Jaynes1}. Nevertheless, all these works presupposed the form of the entropy right from the beginning. A different and axiomatic route was adopted by Shore and Johnson (SJ) \cite{SJ} who considered the MaxEnt as their point of departure and proved that the MaxEnt uniquely yields the probability distribution if self-consistent inferences from the data are to be drawn. Once the position of the BGS entropy is secured in this way, then the MaxEnt procedure is safely used to obtain the concomitant distribution. In this regard, SJ axioms secure both the soundness of the MaxEnt procedure and legitimacy of the probability distribution obtained through MaxEnt \cite{Presse5}. Therefore, any violation of the SJ axioms casts doubt on both the raison d'etre of the MaxEnt and the concomitant probability distribution for the particular entropy measure under scrutiny.

SJ criteria are composed of four axioms \cite{SJ,Presse}. The first one ensures the uniqueness of the maximization solution of the variational functional $H$ written in terms of the set of probabilities $\{ p \}_{k=1}^n $ while the second one warrants the inference to be independent of the chosen coordinate system. The third one, the subset independence, states that whether one treats an independent subset of system states in terms of separate conditional probabilities or fully in terms of joint probabilities should not matter. The last axiom i.e. system independence ensures the absence of biases when the systems are independent in the face of data which do not couple them. In particular, considering two systems $A$ and $B$ with probabilities $\left\lbrace u_{i}\right\rbrace $ and $\left\lbrace v_{j}\right\rbrace $, respectively, bringing them together implies new bins with the joint probability $p_{ij} = u_{i}v_{j}$.

It is worth noting that the word \textit{independence} has different connotations regarding the fourth and third axioms. The former refers to a direct probabilistic context whereas the latter rather implies that the MaxEnt procedure should yield the same results regardless of the number of states one considers. Imagine that, recording the data of an experiment, we do not know whether we have detected all possible states (events) of the phenomenon under scrutiny. However, even when this is the case, the subset independence ensures that the MaxEnt yields correct results regardless of the subset available to us.

The fourth axiom is recently shown to be violated by the nonadditive $ q $-entropy \cite{Tsallis} by Press{\'e} \textit{et al.} \cite{Presse}. In other words, the distributions associated with the $ q $-entropy do not obey the probability multiplication rule and in turn imply biases even when there is no coupling between the constraints imposed by the data. This in turn jeopardizes the validity of the associated $q$-distributions and moreover makes one doubt all the findings in the field succeeded through fitting with data \cite{Presse2}. It is worth noting that this violation cannot be explained away by concepts from statistical mechanics such as extensivity, since SJ axioms ensure consistent inference and are insensitive to the properties of the system \cite{Presse3,Presse4}. SJ axioms have precedence, since it is interested in consistent and unbiased inference i.e. pre-maximization \cite{Presse3}. Skipping this vital step implies also the failure of the post-maximization related arguments such as extensivity but not vice versa.

Before proceeding further, however, it is very important to understand one subtle issue about SJ criteria. These axioms, even before their explicit statements, assume the following structure
\begin{eqnarray}\label{departure}
H \left( \left\lbrace p \right\rbrace  \right) - \lambda \rbr{\sum_{k} p_{k} a_k-\overline{a}}
\end{eqnarray}  
as their point of departure where $\lambda$ and  $\overline{a}$ are Lagrange multiplier and the measured average of the quantity $a$, respectively. Our focus for now, instead of the axioms themselves, is this equation which at first glance seems very intuitive, since it uses the well-known linear averaging scheme at its foundation. All of the SJ axioms follow this structure and yield consistent inference only when this form of averaging is considered. In fact, this choice of averaging scheme is not accidental at all, since there is another fundamental reason for it aside from being intuitive. BGS entropy, in its formulation by Jaynes and Shannon in particular, is constructed from an ingredient called the information gain (or surprise) i.e. $- \ln\left(\,p_{i}\right)$. As a matter of fact, BGS entropy is just the linear average of this information gain      
\begin{eqnarray}\label{shannon-info}
H(\{p\}) = \left\langle - \ln \left(\,p_{i}\right)\right\rangle_{p} = \sum_{i=1}^{n}  p_{i} \left[ - \ln \left(\,p_{i}\right) \right] \,,
\end{eqnarray}  
where the expression $\left\langle \cdot\right\rangle_{p} $ indicates that the linear average is taken over the probability $p_{i}$. The usual recipe to form generalized entropies is either by somehow deforming the information gain or adopting a different averaging scheme (or even both). For example, the R{\'e}nyi entropy preserves the same information gain expression as the BGS entropy but uses an exponential averaging procedure \cite{Masi}. On the other hand, the nonadditive $q$-entropy $H_{q}$ differs than the BGS entropy $H$ in both its information gain expression and averaging procedure: 
\begin{eqnarray}\label{tsallis}
\nonumber
H_{q}(\{p\}) &=& \left\langle - \ln_q\left(\,p_{i}\right)\right\rangle_{\widetilde{P}} = \sum_{i=1}^{n}  \widetilde{P}_{i} \left[ - \ln_{q} \left(\,p_{i}\right) \right]\\
&=& n^{q-1} \sum_{i=1}^{n}  \frac{p_{i}^{q}-p_i}{1-q}\,,
\end{eqnarray}  
where $-\ln_q \left(\,p_{i} \right)$ is the $q$-deformed information gain and $\left\langle \cdot\right\rangle_{\widetilde{P}} $ denotes linear averaging albeit over the distribution $\widetilde{P}_i$ which is equal to $\frac{p_i^q}{\sum_{k=1}^{n}p^q_{k,\mathrm{eq}}}$ (or $n^{q-1} p_i^q$ explicitly) where the subscript ``eq" denotes the uniform probability distribution \cite{Oikonomou}, and the $q$-deformed logarithm reads $\ln_q (x) = \frac{x^{1-q}-1}{1-q}$ (see also Refs. \cite{R1,R2}). If one now compares the BGS and the $ q $-entropy given by Eqs. (\ref{shannon-info}) and (\ref{tsallis}) respectively, two main differences are clear: the $q$-entropy is obtained first by deforming the information gain and then using a linear averaging over a distribution different from the one used in the BGS entropy. Then, it is easy to see how and why the $q$-entropy might fail in satisfying some or all of the SJ criteria which in fact depend on the preceding functional given by Eq. (\ref{departure}) suitably tailored exclusively for the BGS entropy. This is the subtle reason why SJ axioms uniquely yield the BGS entropy. This juxtaposition of two different entropies provides us with the clue on how to proceed, since it now becomes apparent that one should use a linear averaging over the distribution $\widetilde{P}_i$ in any calculation regarding the nonadditive $q$-entropy (exactly as one would do the same for the BGS entropy by relying on the distribution $p_{i}$) as consistency demands \cite{foot}. This crucial observation, as we will see, is enough to remove all the aforementioned obstacles set by the SJ axioms for $H_q$. Lastly, note that we do not impose or define a new averaging scheme for the $q$-entropy but rather make obvious what is already present in their structure leaving no room for any \textit{ad-hoc} manoeuvre.

Before moving on with the explicit calculations, another essential ingredient should be explicated: $H_q$ is mostly written as the average of the $q$-deformed information gain where average is taken over the distribution $p_i^q$, i.e. $\av{-\ln_q(p_i)}_{p^q}=\sum_i \frac{p_i^q-p_i}{1-q}$, instead of the distribution $\frac{p_i^q}{\sum_{k=1}^{n}p^q_{k,\mathrm{eq}}}$ employed above. However, $H_q$ written in terms of averaging over $p_i^q$, although the mostly used one in the literature, simply violates the second SJ axiom \cite{Abenew}. The second axiom ensures the coordinate-independent inferences and requires a consistent discrete/continuum transition. As shown in Ref. \cite{Abenew}, the nonadditive $q$-entropy averaged with $p_i^q$ does not have a consistent continuous counterpart while it has recently been shown that $q$-deformed information gain averaged with $\frac{p_i^q}{\sum_{k=1}^{n}p^q_{k,\mathrm{eq}}}$ leads to a consistent relative entropy expression \cite{Oikonomou}, conforming to the second SJ axiom. Therefore, we do not explicitly consider this axiom here and refer the reader to Refs. \cite{Oikonomou} and \cite{Abenew} instead.

Armed now with the explicit expression of $H_q$ and the associated averaging scheme, we study the system independence axiom by considering, instead of the BGS functional in Eq. (\ref{departure}), the following variation functional
\begin{eqnarray}\label{Eq02}
\Lambda=H_q - \lambda_a \rbr{\sum_{i,j}\widetilde{P}_{ij} a_i-\overline{a}} - \lambda_b \rbr{\sum_{i,j}\widetilde{P}_{ij} b_j-\overline{b}}
\end{eqnarray}
with $H_q=H_q(A+ B)=\sum_{i,j}f(p_{ij})$, where we omitted normalization constraint for simplicity. Then, the variation condition $\delta \Lambda=0$ yields
\begin{eqnarray}\label{eq3c}
(p_{ij})^{1-q}f'(p_{ij})- K_{ij} &=& 0\,,
\end{eqnarray}
where $K_{ij}= \lambda_a N^{(q-1)} q a_i + \lambda_b N^{(q-1)} q b_j$ and $N = n_{A} n_{B}$. 
Assuming the general dependence $p_{ij}=p(u_i,v_j)$ and applying the derivative $\frac{\partial}{\partial u_i}$ on Eq. (\ref{eq3c}) we have
\begin{eqnarray}\label{eq01}
0 &=& \big[(1-q) f'(p_{ij}) + p_{ij} f''(p_{ij})\big] \frac{\partial p_{ij}}{\partial u_i}.
\end{eqnarray}
We then proceed to take the derivative $\frac{\partial}{\partial v_j}$  of Eq. (\ref{eq01}) so that
\begin{eqnarray}\label{eq02}
0 &=& (2-q) f''(p_{ij}) + p_{ij} f'''(p_{ij})\,.
\end{eqnarray}
Solving the differential equation in Eq. (\ref{eq02}) we determine $f(p_{ij})$ as
\begin{eqnarray}\label{fF}
f(p_{ij}) = -C_1\frac{(p_{ij})^q}{q(1-q)}+ C_2\, p_{ij} + C_3.
\end{eqnarray}
Accordingly, we have
\begin{eqnarray}\label{SJEnt}
\nonumber
&&H_q(A+ B) = \sum_{i=1}^{n_A}\sum_{j=1}^{n_B}f(p_{ij})=  N\,C_3+\\ &&+\frac{1}{(1-q)}\sum_{i=1}^{n_A}\sum_{j=1}^{n_B}\bigg[-\frac{C_1}{q}(p_{ij})^{q-1} + (1-q)C_2\bigg]\, p_{ij}.\hskip0.8cm
\end{eqnarray}
Note that the terms $C_{i}$ ($i=1,2,3$) may generally depend on $q$, $n_{A}$ and $n_{B}$, since we consider the composite system i.e. $A+B$. Without loss of generality we set $C_2=1/(q-1)$, $C_3=0$ and substitute $C_{A+ B}\equiv (-C_1/q)^{1/(q-1)}$ with $C_1<0$, so that the composite nonadditive $ q $-entropy reads
\begin{eqnarray}\label{SJEnt2}
H_q(A+ B) = \sum_{i=1}^{n_A}\sum_{j=1}^{n_B} p_{ij}\ln_q\left(\frac{1}{C_{A+ B}\,p_{ij}}\right) \,.
\end{eqnarray}
However, the SJ approach yields the entropy of the composite system $A+ B$ rather than the entropy of a single system. To this aim, one needs to specify both a joint probability composition rule $p_{ij}$ and an entropy composition rule. Only then one can potentially determine the desired single system entropy.\\

In accordance with the fourth SJ axiom, we adopt the multiplicative joint probability composition rule as $p_{ij}=p(u_i,v_j)=u_iv_j$ and assume $C_{A+B}=C_AC_B$, where $C_{A}$ and $C_B$ are components of the composite system $C_{A+ B}$ for each system separately. This joint probability rule is consistent with the formula $N=n_An_B$ already used above. Then, Eq. (\ref{SJEnt2}) can be written as 
\begin{eqnarray}\label{SJEnt3}
\nonumber
H_q(A+ B)= \sum_{i=1}^{n_A} u_{i}\ln_q\left(\frac{1}{C_A\,u_{i}}\right) + \sum_{j=1}^{n_B} v_{j}\ln_q\left(\frac{1}{C_B\,v_{j}}\right) \\
 + (1-q)\sum_{i=1}^{n_A} u_{i}\ln_q\left(\frac{1}{C_A\,u_{i}}\right)\sum_{j=1}^{n_B} v_{j}\ln_q\left(\frac{1}{C_B\,v_{j}}\right)\,.\hskip0.5cm
\end{eqnarray}
Assuming that the composite system $ q $-entropy $H_q(A+ B)$ satisfies the $q$-additivity rule,  
\begin{eqnarray}\label{addProp}
\nonumber
H_q(A+ B)=H_q(A)+H_q(B)+(1-q)H_q(A)H_q(B)\,,\\
\end{eqnarray}
it is easy to see that the following single system $ q $-entropy can be traced out
\begin{eqnarray}
H_{q}(\{p\})=\sum_{i=1}^{n} p_{i}\ln_q\left(\frac{1}{C\,p_{i}}\right).
\end{eqnarray}
Furthermore, since the second axiom should be satisfied by this expression, we finally obtain
\begin{eqnarray}\label{finEnt}
H_{q}(\{p\}) = n^{q-1}\sum_{i=1}^{n} p_{i} \ln_q\left(\frac{1}{p_{i}}\right)
\end{eqnarray}
apart from an additive factor. This is indeed the $ q $-entropy introduced in Eq. (\ref{tsallis}).

As shown above the $q$-entropy in Eq. (\ref{finEnt}) satisfies the system independence and coordinate invariance axioms. Due to the monotonicity of its slope, it also satisfies the uniqueness axiom as well. Thus, the fourth and last axiom we need to explore is the subset independence, which requires \cite{Presse5}
\begin{eqnarray}\label{qesc1}
D_{\ell j k }\left[H_q(\{p\})-\lambda \av{a}_{*}\right]=0\,.
\end{eqnarray} 
The operator $D_{\ell j k }$ is defined as $D_{\ell j k }:=\frac{\partial}{\partial p_{\ell}} \left(\frac{\partial}{\partial p_{j}}-\frac{\partial}{\partial p_{k}}\right) $. Then, due the trace-form feature of $H_q$ we have $D_{\ell jk}H_q=0$ and for $\av{a}_{*} \to \av{a}_{\widetilde{P}} $ one can easily verify that $D_{\ell jk}\av{a}_{\widetilde{P}}=0$. Hence, we proved that the $q$-entropy in Eq. (\ref{finEnt}) with the averaging procedure $\av{\cdot}_{\widetilde{P}}$ satisfy all four SJ axioms, leaving no room for the artificial biases or any other inconsistencies.

One can now answer another important question regarding $H_q$. In this field, one often relies on the use of the so-called escort distributions to calculate the averages. These distributions are of the form $\frac{\sum_{i=1}^{n}p_i^qa_i}{\sum_{k=1}^{n}p_k^q}$. The question then remains as to whether the escort distribution based averaging scheme, substituted in Eq. (\ref{Eq02}), can provide conformity to the SJ axioms warranting a consistent inference. The answer is no and this can be seen first from considering the axiom of the subset independence in Eq. (\ref{qesc1}). Then, for any trace-form entropy, the former equation reduces to
\begin{eqnarray}\label{escort1}
D_{\ell jk}\left[\frac{\sum_{i} p_i^q a_i}{\sum_{i}p_i^q}\right]=0
\end{eqnarray}
yielding after some algebra the following relation
\begin{eqnarray}\label{escort4}
\left(\frac{p_j}{p_k}\right)^{q-1} = \frac{a_\ell+a_k -2 \frac{\sum_{i} p_i^{q} a_i}{\sum_i p_i^q}}{a_\ell+a_j -2 \frac{\sum_{i} p_i^{q} a_i}{\sum_i p_i^q}}\,.
\end{eqnarray}
This violates the subset independence axiom, since the probability of an energy state $a_j$ depends on another energy state $a_{\ell}$ as well. Thus, the escort mean value is not a remedy for any trace-form entropy as far as one considers the SJ axioms. In other words, not only $q$-entropy, but also BGS entropy fails to conform to the subset independence axiom when the escort averaging is used. The SJ criteria heavily rely not only on the definition of the entropy but also on the averaging scheme employed in the functional.

In order to see that the use of the escort distributions does not conform to the system independence axiom regarding the $q$-entropy, consider Eq. (\ref{Eq02}) but with escort averages now so that the functional reads
\begin{eqnarray}\label{Eq02new}
\Lambda=\sum_{i,j}f(p_{ij}) - \sum_{x=a,b}\lambda_x \rbr{\frac{\sum_{i,j}(p_{ij})^q x_i}{\sum_{k,l}(p_{kl})^q}-\overline{x}} .
\end{eqnarray}
After the variation condition i.e. $\delta \Lambda=0$, the above functional implies
\begin{eqnarray}
f'(p_{ij}) =\frac{q(p_{ij})^{q-1}}{\sum_{k,l}(p_{kl})^q} \sum_{x=a,b} \lambda_x\rbr{x_i-\frac{\sum_{i,j}(p_{ij})^q x_i}{\sum_{k,l}(p_{kl})^q}}. \;
\end{eqnarray}
Multiplying the equation above with $p_{ij}$ and summing over all $i,j$'s we have
\begin{eqnarray}
\sum_{k,l} p_{kl} f'(p_{kl})=0\,,
\end{eqnarray}
since $\sum_{k,l}(p_{k,l})^q\neq0$. Note that we changed the dummy variables $(i,j)\to(k,l)$ after the summation.
Taking first the derivative $\frac{\partial}{\partial u_i}$
\begin{eqnarray}
\sum_{l} \left[ f'(p_{il}) + p_{il} f''(p_{il})\right]\frac{\partial p_{il}}{\partial u_i}=0
\end{eqnarray}
and then the derivative $\frac{\partial}{\partial v_j}$, we have
\begin{eqnarray}
\nonumber
0&=& \left[2f''(p_{ij}) + p_{ij}f'''(p_{ij})\right]\frac{\partial p_{ij}}{\partial v_j}\frac{\partial p_{ij}}{\partial u_i} \\
&+& \left[f'(p_{ij}) + p_{ij} f''(p_{ij})\right]\frac{\partial^2 p_{ij}}{\partial v_j\partial u_i}
\end{eqnarray}
Applying again the multiplicative probability composition rule $p_{ij}=u_i v_j$, we finally see that
\begin{eqnarray}
f'(p_{ij}) + 3p_{ij} f''(p_{ij}) + p_{ij}^2 f'''(p_{ij}) =0
\end{eqnarray}
whose solution can be explicitly given as
\begin{eqnarray}
f(p_{ij})=c_1+c_2 \ln(p_{ij}) + \frac{c_3}{2}\ln^2(p_{ij})\,.
\end{eqnarray}
Obviously, $\sum_{i,j}f(p_{ij})$ in this case is not the nonadditive $q$-entropy given in Eq. (\ref{finEnt}).

To conclude, SJ criteria are indeed powerful and cannot be neglected if one seeks an unbiased inference when there is none in the data. Therefore, they cannot be generalized as one would usually do in the field of generalized entropies. However, SJ criteria implicitly assume that the averaging procedure underlying the entropy and data is the ordinary linear average. It is in this sense that SJ singles out the BGS entropy. If one does not change the averaging procedure and try to fulfil these axioms with another entropy measure, one is bound to violate some or all of these axioms, implying lack of coordinate invariance or creating artificial biases, for example. If one wishes to adopt a new entropy measure, the point of departure should be to determine the averaging scheme that will be compatible with it. Only then, the conformity of the new entropy to these axioms can be investigated. The SJ criteria should be understood as singling out the BGS entropy only when linear averages are employed if a consistent inference is the goal. However, even BGS entropy, as we have demonstrated, fails for example to satisfy one of these axioms, namely subset independence, when a different averaging scheme is employed. The BGS entropy is obtained by the linear average over the distribution $p$ of the information gain whereas the non-additive $q$-entropy $H_q$ is formed by the linear average over the distribution $\widetilde{P}$ of the $q$-deformed information gain. As a result, it is only natural to conclude that the SJ criteria are conformed by the BGS entropy when the linear averages over $p$ are employed while the same set of criteria is satisfied by $H_q$ when the linear averages over the distribution $\widetilde{P}$ are used. SJ criteria choose the entropy and the averaging that should be employed in a unique manner so that the mostly employed escort distributions too fail to conform to the SJ axioms as we have shown. Of course, it is true that the linear averaging over the distribution $p$ seems more intuitive, since it is almost a matter of habit now with its historical arsenal. However, this does not negate the need for new entropies and new averaging schemes. It remains to be seen in the future as to why nature, if it ever does so, chooses such multitude of entropies and different schemes of averaging.

\begin{acknowledgments}
	The authors acknowledge the ORAU grant entitled ``Casimir light as a probe of vacuum fluctuation simplification" with PN 17098. T.O. acknowledges the state-targeted program ``Center of Excellence for Fundamental and Applied Physics" (BR05236454) by the Ministry of Education and Science of the Republic of Kazakhstan. 
\end{acknowledgments}



\begin{thebibliography}{0}
\bibitem{foot} One should not confuse $\widetilde{P}_i$ which is equal to $\frac{p_i^q}{\sum_{k=1}^{n}p^q_{k,\mathrm{eq}}}$ with the usually employed escort distributions $\frac{\sum_{i=1}^{n}p_i^q}{\sum_{k=1}^{n}p_k^q}$, since denominators are different. Moreover, note that we explicitly show the violation of the third and fourth SJ axioms when escort averages are employed. We propose to call $\widetilde{P}_i$ as quasi-escort distribution.
\end{thebibliography}
\end{document}